%% file: main.tex
\documentclass[11pt]{article}
\usepackage[preprint]{acl}
\usepackage{times}

\usepackage{latexsym}
\usepackage[T1]{fontenc}
\usepackage[utf8]{inputenc}
\usepackage{microtype}
\usepackage{graphicx}
\usepackage{amsmath,amssymb,booktabs,longtable,array}
\usepackage{url}
\newcommand{\jev}{Jev-Guard}

\title{Jev-Guard: Fast Routing and Auditing Agent Actions over Editable Policy Trees}

\author {
    {\bf Wenbin Hu$^{1}$}\thanks{Equal Contribution},
    {\bf Huihao Jing}$^{1*}$,
    {\bf Haochen Shi$^{1}$},
    {\bf Yuxuan Liu$^{1}$}, 
    {\bf Changxuan Fan$^{1}$}, \\
    {\bf Sirui Zhang$^{1,2}$},
    {\bf Kawai Chung$^{1}$},
    {\bf Hanyu Yang$^{2}$},
    {\bf Zhenhuan Wang$^{2,3}$},
    {\bf Ziyi Chen$^{2}$}, \\
    {\bf Haoran Li$^{1}$}\thanks{Corresponding author},
    {\bf Yangqiu Song$^{1}$}\\
    $^{1}$HKUST, $^{2}$ModeIO, $^{3}$CUHK (Shenzhen) \\
    \texttt{Project Lead: Wenbin Hu, whuak@connet.ust.hk}
}

\begin{document}
\maketitle
\begin{abstract}
Agent harnesses require a low-latency guardrail model together with customizable audit policies. However, deploying an auto-regressive guardrail model remains challenging: such models incur high inference latency and are often poorly aligned with customized policies. To address these problems, we present \jev, a unified model for safety classification and policy routing. We further construct a policy tree over which \jev\ performs routing, spanning diverse domains including content moderation, cybersecurity, privacy, and legal compliance. Specifically, \jev\ first navigates the policy tree to select the applicable audit policy given the caller's audit context, and then performs safety classification conditioned on the retrieved policies. On 7,033 evaluation cases, \jev\ achieves 93.72\% safety accuracy, a mean latency of 58\,ms, and 93.10\% hierarchical routing accuracy. The full routing-and-classification pipeline averages 240\,ms. Furthermore, we integrate \jev\ into the Hermes harness framework. 
Across agent trajectories, its separate permission gate cuts mean review time per task from 4.80\,s for an SFT-LLM judge to 1.33\,s, with no policy violations are observed for either guard.
\end{abstract}
\input{sections/overview}
\input{sections/introduction}
\input{sections/method}
\input{sections/data-construction}
\input{sections/experiments}
\input{sections/discussion}
\clearpage
\input{sections/limitations}
\bibliography{references}
\clearpage
\appendix
\input{sections/app-data}
\input{sections/app-tree}
\input{sections/app-native}

\input{sections/full-scores}

\input{sections/app-edits}

\input{sections/detailed-routing}
\input{sections/diagnostic-cases}
\input{sections/baseline-details}
\input{sections/app-agent}
\input{sections/app-notes}
\end{document}

%% file: sections/overview.tex
\begin{figure*}[t]
  \centering
  \includegraphics[width=\textwidth]{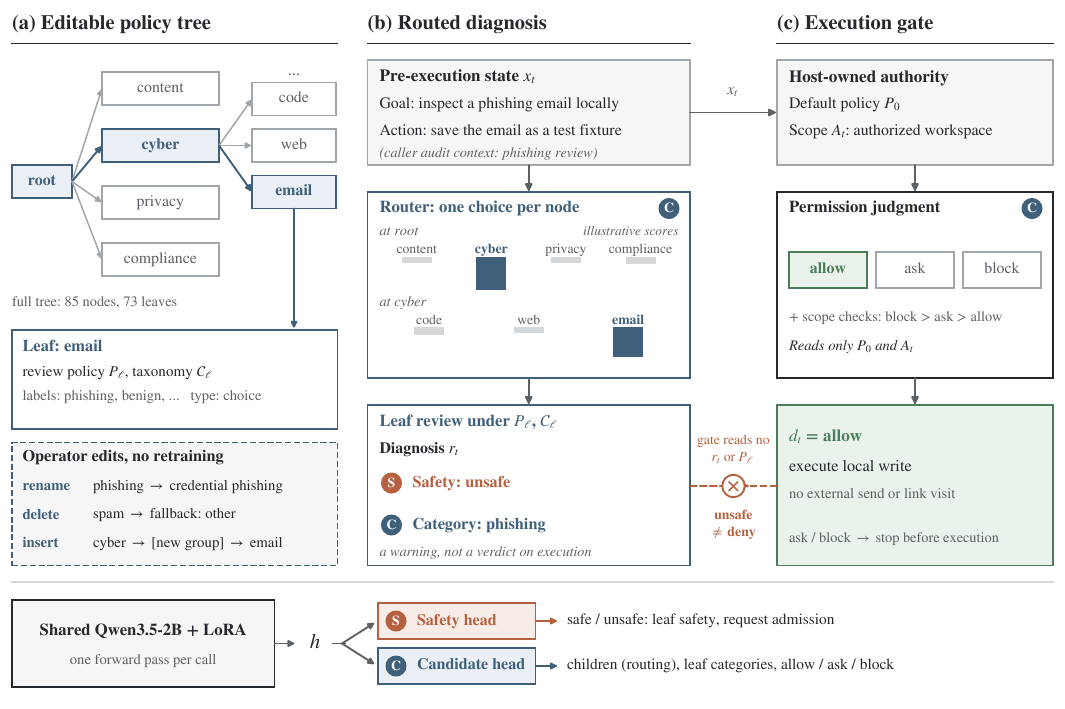}
  \caption{\textbf{Policy tree, routed diagnosis, and execution gate.} (a) The tree is editable without retraining. (b) The router picks one child per node, then the leaf review diagnoses. (c) A separate permission call decides execution without reading the diagnosis. S and C mark the heads. Scores and edits are illustrative.}
  \label{fig:overview}
\end{figure*}

%% file: sections/introduction.tex
\section{Introduction}

Agent harnesses execute tool calls on behalf of users, and every action must be reviewed before it takes effect \citep{wang2026agentspec,mou2026toolsafe}. Such review sits on the critical path of each step, so it demands a low-latency guardrail model. It also demands customizable audit policies: sending a public announcement to an approved partner differs from sending a private document to an address found in a webpage, and the two cases require different audit tasks. Across content moderation, cybersecurity, privacy, and legal compliance, the same agent action may call for vulnerability analysis, a disclosure check, or a review under a particular platform policy. An unsafe label alone identifies neither the audit task nor the execution authority.

Existing guardrail models fall short of these two requirements. First, policy-conditioned safeguards typically generate an auto-regressive rationale before emitting a verdict \citep{li2025gspr,liu2026apt,chen2025shieldagent}, which incurs high inference latency at every tool invocation. Second, such models are often poorly aligned with customized policies. Fixed taxonomies restrict deployments to a shared vocabulary, whereas placing all natural-language policies into the context for every action enlarges the input and introduces irrelevant candidates. Hierarchical labels help organize categories \citep{li2025higuard}, but they do not let operators rename a category, remove it with a fallback, or insert a new routing level without retraining. Such edits change the decision interface even when the underlying action stays fixed, so a frozen model must follow the changed rules.

To address these problems, we present \jev, a unified model for safety classification and policy routing. Following the direct structured-decision paradigm of Jev \citep{typesafejev} and the Open-Jev implementation family \citep{openjev}, \jev\ is a Qwen3.5-2B classifier that replaces generated rationales with direct prediction heads. We further construct a versioned \emph{policy tree} over which \jev\ performs routing, with 85 nodes and 73 leaves spanning content moderation, cybersecurity, privacy, and legal compliance. Given an agent action, \jev\ first navigates the tree through local decisions to select the applicable audit policy given the caller's audit context, and then performs safety classification conditioned on the retrieved leaf rules and taxonomy. Operators can inspect and revise the tree without retraining, and only the policies relevant to the current action enter the context.

We evaluate \jev\ on 7,033 cases. It achieves 93.72\% safety accuracy with a mean latency of 58\,ms and 93.10\% hierarchical routing accuracy. The full routing-and-classification pipeline averages 240\,ms. Beyond classification, we integrate \jev\ into the Hermes harness \citep{hermes}, where an independent permission judgment and scope checks govern execution, and native leaf predictions serve as diagnostics. This separation matters for legitimate analysis of harmful material: recognizing phishing content should not automatically prohibit saving it for an authorized investigation. In agent trajectories built on executable simulated environments \citep{debenedetti2024agentdojo}, the separate permission gate reviews faster than an LLM judge but completes fewer normal tasks, and neither guard shows policy violations.
Our contributions are threefold:
\begin{enumerate}
    \item \textbf{A unified model and editable policy tree.} We present \jev, a 2B direct classifier that performs hierarchical policy routing and policy-conditioned safety classification over an editable policy tree spanning content moderation, cybersecurity, privacy, and legal compliance.
    \item \textbf{A four-domain data construction pipeline.} We map tasks from the four domains to policy leaves while preserving source-specific safety meanings and annotation masks.
    \item \textbf{Strong accuracy at low latency.} On 7,033 evaluation cases, \jev\ achieves 93.72\% safety accuracy and 93.10\% hierarchical routing accuracy at 58\,ms mean latency (240\,ms for the full pipeline), and in Hermes the permission gate reviews faster than an LLM judge at a lower normal-task completion rate.
\end{enumerate}

\section{Related Work}
\paragraph{Policy-conditioned safeguards.}
GSPR trains safeguards to reason across supplied safety policies and taxonomies \citep{li2025gspr}. Augmented policy training studies generalization to unseen policy domains \citep{liu2026apt}. Hi-Guard combines policy-aligned moderation with hierarchical labels \citep{li2025higuard}. These methods establish policy conditioning and hierarchical labels. We instead freeze the model and perturb rules, candidate wording, membership, and tree topology to test whether software edits produce the intended decisions.

\paragraph{Agent runtime protection.}
AgentSpec specifies runtime constraints with triggers, predicates, and enforcement mechanisms \citep{wang2026agentspec}. ShieldAgent evaluates agent behavior using verifiable policy reasoning \citep{chen2025shieldagent}, and ToolSafe studies proactive review at tool invocation steps \citep{mou2026toolsafe}. These systems enforce action policies. We study a direct classifier with a separate audit tree and permission gate, tracing which policy each decision reads. Our rollouts evaluate the integrated gate, whose policy is separate.

\paragraph{Decision models and agent evaluation.}
Jev motivates direct structured decisions rather than autoregressive explanations \citep{typesafejev}. We build on the Open-Jev implementation family \citep{openjev}. We retain its head architecture and add four-domain task mapping and paired audit diagnostics. Official Jev is an external baseline. AgentDojo provides executable simulated environments for testing prompt injection attacks and defenses \citep{debenedetti2024agentdojo}. We use these tools in native Hermes rollouts \citep{hermes}, separating full attacker goals from partial policy violations.

%% file: sections/method.tex
\section{Jev-Guard}
\label{sec:method}

\paragraph{Overview.}
Given a pre-execution state $x_t$, \jev\ (i)~\emph{routes} from the root of the policy tree to one leaf, (ii)~\emph{reviews} the action under that leaf's policy and taxonomy to produce a diagnosis $r_t$, and (iii)~passes it to a separate \emph{execution gate} (Figure~\ref{fig:overview}).

\subsection{State and Policy Interface}
\label{sec:interface}
Let $x_t=(u,a_t,h_t,z_t)$ denote the state available before action $a_t$: the user goal $u$, the proposed tool name and arguments, relevant prior observations $h_t$, and the caller-supplied audit context $z_t$. The host separately supplies the authorized scope $A_t$ and the default policy $P_0$.

The audit context $z_t$ states which audit is requested and, for named-policy branches, which law or provider policy governs. Routing is therefore \emph{caller-conditioned capability dispatch}: the model selects the leaf that matches the supplied intent and policy identity, rather than discovering the relevant policies from the evidence alone. One request selects one leaf, so an action needing several audits requires separate requests.

\subsection{Policy Tree}
\label{sec:policy-construction}
The policy configuration is a rooted tree $\mathcal{T}$. Each internal node stores a description of each child. Each leaf $\ell$ stores a task description $d_\ell$, a local review policy $P_\ell$, a taxonomy $\mathcal{C}_\ell$ with label definitions, and a prediction type (choice, multi-label, or ordinal). Binary safety is optional, since privacy-topic and vulnerability-type leaves need not define it. The evaluated tree has 85 nodes and 73 leaves under four top-level domains: content, cyber, privacy, and compliance. Cyber and privacy branches distinguish audit capabilities (e.g., code vulnerability, phishing, injection), whereas compliance branches distinguish the governing law or provider policy.

The next section describes how the tree is constructed, and the appendices list every leaf.

\subsection{Shared Classifier for Routing and Review}
\label{sec:model}
We adapt the language component of Qwen3.5-2B \citep{qwen35} with LoRA \citep{hu2022lora} and two task heads \citep{typesafejev,openjev}, shared by routing and review. For a serialized policy-and-evidence input $q$, let $h=f_\theta(q)$ be the last unmasked hidden state.

\paragraph{Safety head.}
A binary head predicts
\begin{equation}
 p_{\mathrm{s}}(y\mid q)=\operatorname{softmax}(W_{\mathrm{s}}h+b_{\mathrm{s}})_y,
\end{equation}
where $y\in\{\text{safe},\text{unsafe}\}$.

\paragraph{Candidate head.}
To support label sets that change across leaves and edits, the candidate head scores each candidate $c$ by the similarity between $h$ and the embedding of its label text:
\begin{align}
 v&=\operatorname{norm}(W_{\mathrm{c}}h),\\
 e(c)&=\operatorname{norm}\!\left(\frac{1}{|\tau(c)|}\sum_{j\in\tau(c)}E_j\right),\\
 s_c&=\gamma\,v^\top e(c).
\end{align}
Here $\tau(c)$ is the token sequence of $c$, $E$ is the frozen input embedding matrix, $\operatorname{norm}$ is $\ell_2$ normalization, and $\gamma$ is a learned positive scale. The embeddings $e(c)$ are frozen and cached, so changing the candidate list needs no new output units. A \emph{choice} task takes the largest softmax score. A multi-label task returns every candidate with sigmoid score at least 0.5, possibly none. An auxiliary ordinal head handles ordered severity labels. Because $e(c)$ depends on label text, rewording a candidate can still change predictions.

\paragraph{Routing and review.}
At an internal node $n$, the router applies the candidate head over the children of $n$:
\begin{equation}
 n' = \arg\max_{c\in\operatorname{children}(n)}
 p_{\mathrm{choice}}(c\mid q(x_t,n)),
\end{equation}
where $q(x_t,n)$ serializes the evidence, caller intent, child descriptions, and the governing policy when required. Routing repeats until a leaf $\ell$ is reached. The review input then combines $d_\ell$, $P_0$, $P_\ell$, $\mathcal{C}_\ell$, and the evidence, and the two heads output $r_t$. A routed review thus needs one forward pass per tree level plus one review pass.

\subsection{Execution Gate}
\label{sec:permission}
\paragraph{Design.}
The diagnosis $r_t$ and the execution decision are separate outputs. Harmful material may legitimately be inspected, for example when phishing content is saved for an authorized investigation, so an \emph{unsafe} diagnosis is a warning, not an automatic denial. The gate therefore reads neither $r_t$ nor the selected leaf policy $P_\ell$, so editing a leaf rule changes the diagnosis but not the gate.

\paragraph{Permission judgment.}
For a tool action, with $\mathcal{D}=\{\text{allow},\text{ask},\text{block}\}$,
\begin{align}
 s_t &= f_{\theta}^{\mathrm{choice}}(x_t,A_t,P_0;\mathcal{D}),\\
 d_t &= \operatorname{combine}(s_t,\operatorname{scope}(x_t,A_t)).
\end{align}
The call reuses the frozen candidate head, with no permission-specific fine-tuning. An allow score below 0.55 becomes ask, and explicit scope rules take precedence ($\text{block}>\text{ask}>\text{allow}$). In automated runs both ask and block stop the action. Request admission applies the safety head to the user intent under $P_0$.

%% file: sections/data-construction.tex
\section{Data and Policy Tree Construction}
\label{sec:data}
This section describes how we build the audit data and the policy tree used to train \jev.

\subsection{Data Collection and Construction}
\paragraph{What we collect.}
We gather public datasets for the four audit domains, and each dataset poses a different judgment task. Content tasks ask whether a prompt or response, in text or image form, is harmful and which harm category applies \citep{ghosh2024aegis,zong2024vlguard,zhu2026safemt}. Cyber tasks include detecting vulnerable code \citep{ding2024primevul,chen2023diversevul}, phishing and malicious web requests \citep{dalton2025phreshphish,liu2024cyberbench}, and prompt injection \citep{yi2025bipia}, classifying the root cause of a reported vulnerability \citep{alam2024ctibench}, and judging whether a request is authorized \citep{wang2026cybergym}. Privacy tasks annotate the practices in a privacy policy \citep{wilson2016opp115,ahmad2021policyie,ravichander2019privacyqa} and judge whether an information flow is appropriate \citep{li2025privaci}. Compliance tasks ask whether a text violates a given law or platform policy, such as GDPR or a provider's terms of service. Most come from OmniCompliance-100K (Omni) \citep{hu2026omnicompliance}, and the rest detect unfair clauses \citep{lippi2019claudette,chalkidis2022lexglue}. The appendix lists every dataset.

\paragraph{Unified records.}
Because the datasets use different formats, we convert every example into one record. It holds the evidence to review, the audit task, the applicable policy, the candidate labels with definitions, the labels, and the policy-tree leaf the example belongs to. The training set has 377,930 records. We remove instructions that ask for generated text, since the model outputs labels directly.

\paragraph{Label provenance.}
Labels differ in reliability, so we record where each comes from. \emph{Original} labels are taken directly from the source dataset. \emph{Mapped} labels are converted from source metadata by an explicit rule, for example an attack-family field mapped to an injection category. \emph{Weak} labels are guessed by simple rules such as keyword matching, for example a privacy topic inferred from a sentence. Weak labels are used in training with a small loss weight of 0.25 and never in evaluation. A category that a source never annotated is treated as unknown rather than negative, so the model is not taught that an unlabeled category is absent.

\input{sections/baseline-table}

\subsection{Policy Tree Construction}
We build the tree manually from source information, not from model predictions, and the model never sees dataset names.

\paragraph{Capability assignment.}
We first assign each dataset an audit capability. Complete functions map to code review, vulnerability descriptions with Common Weakness Enumeration (CWE) labels \citep{mitrecwe} to root-cause classification, severity ratings to an ordinal task, and emails, URLs, and HTTP traffic to their own cyber leaves. Datasets with compatible tasks share a leaf. Compliance datasets keep the specific law or provider policy as their branch, grouped by provider and by law. The result has four top-level domains, 85 nodes, and 73 leaves.

\paragraph{Leaf taxonomy.}
Each leaf has its own list of candidate labels, rewritten into readable names. CWE codes are mapped to 20 defect families (e.g., CWE-119 and CWE-787 to \emph{Buffer bounds}), and broad codes that name no specific defect are masked. For policy clauses, we remove duplicated headings and numbers and keep distinct clause details as definitions. Each leaf also records the safety meaning of its native labels, since not every task has a safe or unsafe label.

\paragraph{Routing examples.}
Routing walks the tree one step at a time, so each record's leaf path gives one training example per branching node, in which the model chooses the next child. We sample 40/30/20/10\% of examples at depths zero through three. The router set has 10,000 records, split 8,000/1,000/1,000 without shared evidence or source groups.

%% file: sections/baseline-table.tex
\begin{table*}[t]
\centering\small
\begin{tabular}{lrrrrrr}
\toprule
Model & Content & Cyber & Privacy & Compliance & Overall & Latency (ms) \\
\midrule
Matched test cases & 2,180 & 3,694 & 655 & 504 & 7,033 & --- \\
\midrule
Official Jev 1.13 (API) & 69.08 & 82.27 & 80.46 & 52.58 & 75.89 & --- \\
Qwen3.5-2B SFT & 99.91 & 89.52 & 91.45 & 95.63 & 93.36 & 513.94 \\
\textbf{Jev-Guard (ours)} & 97.29 & 91.66 & 91.60 & 96.03 & 93.72 & 58.31 \\
\bottomrule
\end{tabular}
\caption{\textbf{Four-domain baseline comparison: safety accuracy (\%).} The first row gives matched cases per domain, and the other rows are the systems compared on the same 7,033 text records. Latency is the local per-case median without routing or permission checks. Prompts and training histories differ across systems.}
\label{tab:baselines}
\end{table*}

%% file: sections/experiments.tex
\section{Experiments}
\subsection{Setup}
\paragraph{Research questions.}
We organize the experiments around four questions. \textbf{RQ1}: Is \jev\ accurate and fast enough to review every action? \textbf{RQ2}: Does a frozen model follow edits to rules, categories, and tree structure? \textbf{RQ3}: What does the policy tree add over presenting all policies at once? \textbf{RQ4}: How does the execution gate compare with an LLM judge?

\paragraph{Training.}
The evaluated checkpoint continues an earlier version for one epoch and is selected at update 1,508. We train rank-8 LoRA and the prediction heads with learning rates $2\!\times\!10^{-5}$ and $5\!\times\!10^{-5}$, respectively, a global batch of 256, and a 16,384-token cap. Selection averages routing, safety, and category scores on 2,048 native and 1,000 router validation records. Reported test safety uses ordinary accuracy.

\paragraph{Evaluation.}
We evaluate one frozen checkpoint as an empirical system study. Native scores describe recorded cohorts, not generalization to unseen policies, and edit experiments measure paired responses to newly specified variants.

\subsection{RQ1: Is \jev\ Accurate and Fast?}
\label{sec:domain-results}
We compare \jev\ with a generative SFT model and official Jev on matched records, evaluate native classification across all four domains, and time the full review pipeline.

\paragraph{Finding 1: Accuracy matches SFT at much lower latency.}
On 7,033 matched records, \jev\ reaches 93.72\% safety accuracy, against 93.36\% for SFT and 75.89\% for official Jev \citep{typesafejev}, and Table~\ref{tab:baselines} gives the breakdown by domain. The 0.36-point gap over SFT has a 95\% interval that includes zero, so we claim no accuracy advantage. The SFT median latency is 8.81 times that of \jev. Prompts and training histories differ across systems, so the comparison does not isolate architecture.

\paragraph{Finding 2: Native accuracy varies widely across tasks.}
\input{sections/domain-detail}
We evaluate the checkpoint on 147,643 native task-case records across 109 benchmark views, keeping the original prompts and labels. Safety accuracy is 91.78\% over the 79,987 records with safety labels, and Table~\ref{tab:domains-main} shows how much it varies across domains and tasks. PrimeVul \citep{ding2024primevul} safety accuracy is 96.86\% on single functions but 50.00\% on function pairs, and root-cause category F1 on CTI \citep{alam2024ctibench} remains low. Overlapping views are not independent datasets.

\paragraph{Finding 3: A full review takes about 240 ms.}
We time 200 unique pre-execution states recorded from agent runs, 30 user requests and 170 tool actions, on one A800 80GB GPU with batch size one, bfloat16/SDPA inference, and eight warmup reviews. In-process timing covers serialization, tokenization, routing, leaf prediction, the independent permission judgment, and explicit rules, and tools are not executed again. Table~\ref{tab:latency} shows that complete review averages 239.64\,ms, of which the permission call takes 74.50\,ms. Over HTTP the average is 249.85\,ms.

\begin{table}[t]
\centering
\small
\begin{tabular}{lrrr}
\toprule
 & Mean & p50 & p95 \\
\midrule
\multicolumn{4}{l}{\emph{Per stage, in-process}} \\
Routing & 87.85 & 72.34 & 153.13 \\
Leaf review & 71.32 & 75.67 & 116.86 \\
Permission judgment & 74.50 & 71.29 & 104.95 \\
Explicit rules & 0.01 & 0.00 & 0.01 \\
\midrule
\multicolumn{4}{l}{\emph{End to end}} \\
Complete, in-process & 239.64 & 228.27 & 352.89 \\
Complete, HTTP & 249.85 & 242.98 & 397.37 \\
\bottomrule
\end{tabular}
\caption{Review latency in ms. Rows are review stages and end-to-end totals, with quantiles computed per stage. In-process rows use 200 states, HTTP uses 647.}
\label{tab:latency}
\end{table}

\subsection{RQ2: Does a Frozen Model Follow Policy Edits?}
\label{sec:edits}
We keep the checkpoint frozen, edit rules, candidate lists, and tree structure, and score only newly specified variants. Table~\ref{tab:edits} summarizes all three edit types.

\begin{table}[t]
\centering
\small
\begin{tabular}{lrr}
\toprule
Condition & Correct & Acc. (\%) \\
\midrule
\multicolumn{3}{l}{\emph{Rule edits, 30 targets and 30 controls}} \\
Choice head, targets & 21/30 & 70.00 \\
Choice head, controls & 30/30 & 100.00 \\
Safety head, targets & 30/30 & 100.00 \\
Safety head, controls & 24/30 & 80.00 \\
\midrule
\multicolumn{3}{l}{\emph{Category edits, 24 actions}} \\
Original categories & 23/24 & 95.83 \\
Rename & 20/24 & 83.33 \\
Add a category & 24/24 & 100.00 \\
Delete with fallback & 12/24 & 50.00 \\
Split & 20/24 & 83.33 \\
\hspace{1em}split-dependent actions only & 0/4 & 0.00 \\
Merge & 16/24 & 66.67 \\
Reverse candidate order & 18/24 & 75.00 \\
\midrule
\multicolumn{3}{l}{\emph{Tree edits, 200 cases}} \\
Original tree & 190/200 & 95.00 \\
Reverse sibling order & 190/200 & 95.00 \\
Move injection leaf to root & 190/200 & 95.00 \\
Insert intermediate group & 172/200 & 86.00 \\
\hspace{1em}cyber only, original tree & 40/50 & 80.00 \\
\hspace{1em}cyber only, group inserted & 22/50 & 44.00 \\
\bottomrule
\end{tabular}
\caption{Frozen-model edit results. Rows are edit conditions or scored heads, and indented rows restrict scoring to affected cases.}
\label{tab:edits}
\end{table}

\paragraph{Finding 4: For rule edits, the two output heads disagree.}
We test five rule families: recipients, publication, storage region, retention, and security authorization. Each has six targets whose gold label changes under the edited rule and six unchanged controls, with gold set by predefined predicates. We score the compliant/violation choice and the binary safety output separately, without scope rules or execution. The choice head is correct on both versions of 21 of the 30 targets and on all 30 controls. The safety head is correct on all 30 targets but only 24 controls, because six legal 14-day retention actions stay unsafe. Tightening the retention limit from 90 to 30 days leaves the choice prediction at compliant for all six targets, although the safety head flips correctly. Head agreement and control correctness therefore need separate checks.

\paragraph{Finding 5: Category edits fail when a new distinction is required.}
We change candidate names, membership, granularity, or order for 24 actions across six families. Deleting a category with an explicit fallback cuts accuracy from 95.83\% to 50.00\%, and reversing candidate order lowers it to 75.00\% even though label meanings are unchanged. Splitting a category scores 83.33\% overall, but all four actions that need the new subtypes fail. For example, reading \texttt{passwords\_0.json} should change from \emph{Credential access} to \emph{Password file access}, but the model predicts \emph{Benign workspace action}. The 20 unaffected cases hide these four failures.

\paragraph{Finding 6: Rearranging the tree is safe, adding a routing level is not.}
We sample ten text cases from each of 20 source-task test splits, giving 50 cases per domain and excluding the 48 cases from an earlier pilot. The source datasets are not held out from training. The original tree routes 190 of 200 cases correctly. Reversing sibling order or moving the injection leaf to the root preserves all 200 predictions. Inserting a semantic intermediate group lowers accuracy to 86\%, and all new errors are in cyber, which falls from 40 to 22 of 50 while the other domains stay at 50 of 50.

\subsection{RQ3: What Does the Policy Tree Add?}
\label{sec:ablation}
We compare hierarchical and flat leaf selection, and then four ways of presenting policies to the reviewer.

\paragraph{Finding 7: Hierarchical routing is not more accurate than flat selection.}
We use 1,000 node-level test records from 977 unique source cases. Each record restarts at the root, both conditions receive the caller audit context and governing policy, and the final leaf is scored. The hierarchy chooses children recursively, while the flat condition selects among the same 73 leaf descriptions in one step. The hierarchy reaches 93.10\% against 92.40\% for flat selection. The 0.70-point gap has a 95\% interval from $-0.10$ to 1.60 under a paired source-cluster bootstrap \citep{efron1994bootstrap}, so we find no conclusive evidence that the hierarchy improves routing accuracy. Of the 69 hierarchy errors, 61 first diverge within cyber, where accuracy is 80.45\%. Because 488 of the 1,000 records are named-policy compliance tasks, these scores measure caller-conditioned dispatch.

\begin{table}[t]
\centering
\footnotesize
\setlength{\tabcolsep}{2.5pt}
\begin{tabular}{lrrrrr}
\toprule
Condition & Safety & Cat. F1 & Strict & Unsupp. & ms \\
\midrule
Tree & 66.28 & 24.73 & 19.66 & 12/128 & 205.22 \\
Flat leaf & 68.60 & 24.44 & 18.80 & 15/128 & 152.20 \\
Oracle leaf & 67.44 & 27.86 & 23.08 & 0/128 & 73.66 \\
Full flat & 79.07 & 20.65 & 2.56 & 115/128 & 725.08 \\
\bottomrule
\end{tabular}
\caption{Policy presentation on 128 cases at 32K. Rows are the conditions in the text. Cat. F1 is observed category F1, Unsupp. is unsupported outputs.}
\label{tab:flat}
\end{table}

\paragraph{Finding 8: The tree is faster and more faithful to the taxonomy than a full flat prompt, but not more accurate.}
A flat prompt holding every leaf's policy and taxonomy is far longer than the deployed 16K context. Only 23 of the 1,000 audit inputs fit, and lengths range from 12,742 to 27,318 tokens. To compare presentations fairly, we raise the cap to 32K without truncation and run four conditions on 128 text cases. \emph{Tree} routes hierarchically and then reviews the selected leaf. \emph{Flat leaf} selects one leaf in one step and uses the same review. \emph{Oracle leaf} reviews with the gold leaf, which a deployed reviewer would not have. \emph{Full flat} places all 73 leaves in one prompt. The cases come from the router test split without images, and inference runs in process because the deployed HTTP interface limits candidates to 512. Table~\ref{tab:flat} reports the results, with safety scored on 86 labeled cases and category metrics on 117. Safety uses the native head even on wrong routes, a missing leaf counts as a failure, and times include routing and leaf review but not permission judgment. Observed category F1 scores only the slots the source annotated. Strict set accuracy additionally requires correct routing and no category outside the native taxonomy, and such categories count as unsupported outputs.

The tree is 3.53 times faster than full flat but slower than one-step leaf selection. Its category F1 is higher than full flat's, while full flat has higher safety accuracy. The paired safety gap is $-12.79$ points, and its exact paired-bootstrap 95\% interval runs from $-25.58$ to 0.00. Of 50 unsafe cases, tree review predicts safe on 27 and produces no leaf on one, against 11 safe predictions for full flat. Of 36 safe cases, tree and full flat have one and seven false positives. Full flat outputs categories outside the native taxonomy on 115 of 128 cases, against 12 for the tree. This measures adherence to the local vocabulary, not semantic errors, and candidate and context changes prevent isolating the effect of tree structure.

\paragraph{Finding 9: Correct dispatch does not guarantee correct diagnosis.}
In compliance, routing is correct on all 41 cases, yet tree, flat-leaf, and oracle review each get only 18 of 41 safety labels right, against 33 for full flat. Privacy oracle exact-set correctness is 0 of 40. These results place the failures after dispatch without identifying their cause.

\subsection{RQ4: How Does the Execution Gate Compare with an LLM Judge?}
\label{sec:e2e}
Hermes \citep{hermes} runs AgentDojo \citep{debenedetti2024agentdojo} tools for workspace, banking, travel, and Slack, with 16 normal and 16 injected tasks per suite. We compare two guards, each combined with explicit scope rules: the \jev\ permission judgment and a separate LLM permission call. Both see identical policy, scope, task, tools, attacks, and budget, and the actor and judge are both \texttt{deepseek-flash} with temperature zero. The full run also has no-guard and rules-only reference arms, 512 episodes in total. We report normal completion, official complete attacker goals, and a supplementary checker for partial policy violations. The checker inspects executed prefixes for unauthorized transfers, recipients, or deletions and covers 48 of 64 attack tasks, leaving 16 unknown. It was frozen after the rollouts, so it was not preregistered, and the 16 uncovered attacks are not counted as safe.

\paragraph{Finding 10: Neither guard shows a violation, but the runs cannot credit either one.}
Both guards have zero covered policy violations and zero complete attacker goals. The unguarded baseline also has none, so these zeros cannot be credited to either guard. The upper end of the 95\% Wilson interval for 0 of 48 is 7.41\%, for this covered sample only.

\paragraph{Finding 11: \jev\ reviews faster than the LLM judge but completes fewer normal tasks.}
\jev\ completes 51 of 64 normal tasks and the LLM judge 57, or 79.69\% against 89.06\%. The 95\% intervals of the two completion rates overlap, so we do not claim a reliable difference. \jev\ is more conservative. It declines one normal user request and 25 proposed actions across ten normal tasks, against one request and three actions across three tasks for the LLM judge. Without per-action gold labels we cannot call these denials false positives, since a task that is normal overall can still contain a mistaken action. In return, review is cheaper. Mean review time per normal task is 1.33 seconds for \jev\ and 4.80 seconds for the LLM judge, and on the 48 normal tasks completed in every condition, mean task time is 7.60 seconds against 10.60 seconds.

%% file: sections/domain-detail.tex
\begin{table*}[t]
\centering\small
\begin{tabular}{llrrrrr}
\toprule
Domain / view & Evaluation task & $N$ & $N_s$ & $N_c$ & Safety acc. & Category F1 \\
\midrule
\textbf{Content} & \textbf{All native views} & 11,909 & 11,909 & 9,806 & \textbf{85.20} & \textbf{61.86} \\
 & AEGIS & 1,679 & 1,679 & 0 & 84.69 & --- \\
 & SafeMT & 997 & 997 & 997 & 100.00 & 84.35 \\
 & Nemotron-3.5 & 1,494 & 1,494 & 1,489 & 56.02 & 36.87 \\
 & VLGuard & 925 & 925 & 925 & 43.03 & 24.97 \\
\midrule
\textbf{Cyber} & \textbf{All native views} & 61,972 & 57,478 & 10,164 & \textbf{93.77} & \textbf{46.96} \\
 & PrimeVul / functions & 24,777 & 24,777 & 545 & 96.86 & 20.45 \\
 & PrimeVul / paired view & 862 & 862 & 431 & 50.00 & 20.26 \\
 & DiverseVul & 15,677 & 15,677 & 0 & 90.20 & --- \\
 & CTI-RCM & 987 & 0 & 987 & --- & 35.36 \\
 & PhreshPhish & 6,236 & 6,236 & 0 & 94.58 & --- \\
 & HTTP anomaly & 1,222 & 1,222 & 1,222 & 83.14 & 83.80 \\
 & Attack technique & 1,088 & 0 & 1,088 & --- & 83.27 \\
 & CyberGym-derived authorization & 2,966 & 2,966 & 0 & 98.85 & --- \\
\midrule
\textbf{Privacy} & \textbf{All native views} & 69,051 & 5,889 & 68,318 & \textbf{82.92} & \textbf{78.40} \\
 & OPP-115 & 697 & 0 & 697 & --- & 71.82 \\
 & PolicyIE-A & 918 & 0 & 918 & --- & 87.36 \\
 & PrivacyQA & 61,157 & 0 & 61,157 & --- & 87.62 \\
 & PrivaCI / GDPR & 3,105 & 3,105 & 3,096 & 89.57 & 58.76 \\
 & PrivaCI / HIPAA & 107 & 107 & 107 & 39.25 & 30.71 \\
\midrule
\textbf{Compliance} & \textbf{All native views} & 4,711 & 4,711 & 4,707 & \textbf{95.31} & \textbf{58.17} \\
 & Unfair-ToS & 1,600 & 1,600 & 1,600 & 96.06 & 75.97 \\
 & Omni / GDPR & 359 & 359 & 359 & 94.99 & 72.98 \\
 & Omni / GitHub terms & 268 & 268 & 268 & 89.93 & 41.79 \\
\bottomrule
\end{tabular}
\caption{\textbf{Native classification by domain and task.} Bold rows are domains covering all their native views, and indented rows are selected views, with all 109 in the appendix. $N_s$/$N_c$ count safety/category supervision, dashes mean missing supervision, and category F1 is micro-averaged.}
\label{tab:domains-main}
\end{table*}

%% file: sections/discussion.tex
\section{Discussion}
\paragraph{Editability is a behavior to test.}
Accepting a new label or node does not ensure correct interpretation. Split and retention probes require checking affected cases, controls, and separate heads. Hierarchy organizes policies, but maintenance benefits remain unmeasured.

\paragraph{Validate diagnosis and permissions.}
Perfect compliance dispatch coexists with poor leaf judgments, including under oracle routing. The permission gate separately reads default policy and scope. Binding tree rules to execution requires a new interface and execution tests. The current results instead support three practical checks: score newly affected cases, retain controls, and measure complete review with normal-task completion.

\paragraph{Report the audit contract.}
Comparisons require explicit audit contracts: caller information, observed labels, and outputs that control execution. An extra category outside a source vocabulary indicates a contract mismatch. Unannotated categories are not automatically semantic errors. 

\section{Conclusion}
\jev\ enables empirical tests of editable audits in four domains. Native classification has lower archived latency than SFT without a conclusive accuracy advantage. Paired tests expose edit failures and diagnostic errors after correct routing. Flat dispatch remains competitive. Full review averages 240\,ms. In agent runs, the gate reviews faster than an LLM judge at a lower completion rate, without established safety gains.

%% file: sections/limitations.tex
\section*{Limitations}
We evaluate one checkpoint, a limited set of policy edits, and one actor API alias. The controlled edit cases are constructed from explicit predicates, not independently annotated production logs. Five rule families and six category families provide limited statistical power. The 200-case tree study is balanced across selected sources and does not estimate deployment prevalence. Confidence intervals over task or source samples do not include training-seed variation or future API model changes.

The full-policy comparison uses 32K inputs beyond the 16K fine-tuning cap and bypasses the deployed 512-candidate HTTP limit. It establishes a matched experimental comparison, not a drop-in deployment configuration. Native and shared-taxonomy evaluations differ in prompts, candidates, and supervision. Unknown category slots are masked. Strict unsupported-output metrics do not establish that all unsupported predictions are semantically incorrect.

The agent experiments use static attacks, simulated tools, one rollout per task, and only partial violation coverage. Sixteen attacks have no reliable partial-policy checker and remain unknown. The supplementary checker was frozen before rescoring, not before the original actor rollouts. Zero baseline violations preclude a measured incremental safety claim. No human-in-the-loop resolution of ask decisions is evaluated. The current leaf policy is not bound to the permission head, so this implementation does not establish end-to-end enforcement of arbitrary tree edits.

Current v3.2 split audits found no overlap under case-ID, source-group, and content checks for the audited train/router splits. They do not fully exclude exposure through all historical v1--v3.1 training data or foundation-model pretraining. Public benchmark familiarity remains possible. We do not claim that the entire source mixture can be redistributed under one license, nor that heterogeneous native safety labels share a universal semantic meaning.

\section*{Ethical Considerations}
The system is a research guard, not a guarantee of safe execution or a replacement for explicit authorization controls. Incorrect denials can impede legitimate work. Incorrect allows can expose sensitive assets. Our action experiments use isolated simulated environments and synthetic boundary-workflow assets, without sending real sensitive data or operating production resources. Evidence from tool outputs is kept distinct from host-owned policy and authorization. Logs used for real deployments would require data minimization and access control, which this study does not validate at production scale.

The training mixture includes harmful text and vulnerable source code for classification. Redistribution must respect each source's access and licensing conditions. Our reports avoid publishing credentials, production traces, or personal user content. Automated coding and drafting tools assisted the implementation and manuscript preparation. The substantive claims here are tied to recorded experiments and source inspection rather than generated hypothetical results.

%% file: sections/app-data.tex
\section{Training and Data Summary}
\label{app:data}
\paragraph{Training.}
The evaluated checkpoint is a LoRA adaptation of Qwen3.5-2B with two prediction heads, frozen during all evaluation. It continues an earlier checkpoint for one epoch, so it is not a fresh base-model run and is not evidence of training to a validation plateau. Model selection uses a fixed validation panel, while reported test safety accuracy is the ordinary fraction correct over labeled cases.

\paragraph{Data composition.}
The training package contains 377,930 native audit records and 8,000 router records. A taxonomy annotation without a valid safety label contributes only to its supported task, and governing policies are part of the input for policy-dependent classification.

\begin{table}[t]
\centering
\small
\begin{tabular}{lrr}
\toprule
Domain & Native train & Validation library \\
\midrule
Content & 212,773 & 1,003 \\
Cyber & 74,193 & 43,764 \\
Privacy & 29,581 & 16,954 \\
Compliance & 61,383 & 3,809 \\
\midrule
Total & 377,930 & 65,530 \\
\bottomrule
\end{tabular}
\caption{Native records by domain, excluding router examples.}
\label{tab:train}
\end{table}

\paragraph{Annotation coverage.}
Taxonomy coverage exceeds annotation coverage. We identify 172 leaf-local category entries across 41 leaves with no positive annotations, so these entries are representable by the candidate interface but their supervised capability is not established. Across all native splits, 4,506 records involve weak rule annotations and 6,954 involve source-label crosswalks. These are not additional training records.

\paragraph{Router set and exposure.}
The router collection has 10,000 examples split 8,000/1,000/1,000. Its test split is not domain-balanced, with 100 content, 312 cyber, 100 privacy, and 488 compliance records, so its overall accuracy is interpreted differently from the balanced 200-case tree-edit study. The checkpoint was developed through several data versions whose sources, sampling, prompts, and taxonomies changed together. The final scores therefore do not isolate the contribution of individual data sources, and earlier training stages may have exposed the model to benchmark data.

%% file: sections/app-tree.tex
\section{Policy Tree Summary}
\label{app:construction}
\label{app:catalog}
The implementation contains three related but distinct kinds of policy. A \emph{native task policy} defines the source classification problem, for example whether an email is phishing or whether a clause violates a platform rule. A \emph{leaf review policy} tells the reviewer how to apply a capability and its taxonomy to the evidence. The \emph{host authorization policy} defines permissible actions and scope in the agent harness. Their labels differ: vulnerable code need not be unauthorized input, and a content label is not an execution permit.

The tree is a manually specified, inspectable interface compiled from source metadata, not a hierarchy learned from test labels. A named-policy branch cannot obtain its governing rule by inferring a legal topic. Its caller context must identify the policy, and the selected policy name must match a supported name. Table~\ref{tab:tree-summary} summarizes the evaluated configuration, which has 85 nodes and 73 leaves.

\begin{table}[t]
\centering
\small
\begin{tabular}{lrl}
\toprule
Domain & Leaves & Prediction types \\
\midrule
Content & 1 & multi-label \\
Privacy & 1 & multi-label \\
Cyber & 10 & choice, multi-label, ordinal \\
Compliance & 61 & choice \\
\midrule
Total & 73 & \\
\bottomrule
\end{tabular}
\caption{Leaves of the evaluated policy tree by domain. Each compliance leaf corresponds to one law or provider policy.}
\label{tab:tree-summary}
\end{table}

%% file: sections/app-native.tex
\section{Native Classification Results}
\label{app:native}
The earlier native test package contains 147,643 task-case records and 109 benchmark views. Views may overlap and are not 109 independent datasets. This evaluation preserves the original test candidates and gold labels. It therefore measures compatibility with the original native interfaces, not performance through the shared policy tree. Table~\ref{tab:native} reports domain-level ordinary safety accuracy and category micro-F1.

\begin{table*}[t]
\centering
\small
\begin{tabular}{lrrrrrr}
\toprule
Domain & Task cases & Safety cases & Safety acc. & Category F1 & Safe $\to$ unsafe & Unsafe $\to$ safe \\
\midrule
Content & 11,909 & 11,909 & 85.20 & 61.86 & 54.31 & 1.76 \\
Cyber & 61,972 & 57,478 & 93.77 & 46.96 & 3.19 & 21.17 \\
Privacy & 69,051 & 5,889 & 82.92 & 78.40 & 28.08 & 10.87 \\
Compliance & 4,711 & 4,711 & 95.31 & 58.17 & 2.48 & 7.04 \\
\bottomrule
\end{tabular}
\caption{Native evaluation in the original test-label space. Scores and conditional error rates are percentages. Category F1 is micro-averaged within each domain. Category-labeled counts are 9,806 (content), 10,164 (cyber), 68,318 (privacy), and 4,707 (compliance). No common-category aggregate is implied.}
\label{tab:native}
\end{table*}

Overall safety accuracy is 73,415/79,987 (91.78\%) on cases with safety supervision. It does not use all 147,643 records as its denominator. Content safety accuracy is accompanied by a 54.31\% conditional false-positive rate on safe examples, illustrating how class prevalence can hide an important weakness. Privacy has many topic-only cases, so its safety denominator is much smaller than its total count. Ordinal severity accuracy is 15.28\% on 1,466 cases, with normalized mean absolute error 0.17271. No strong severity capability is established.

The earlier routing report gives 93.20\%, while the supplemental paired tree/flat protocol gives 93.10\%. The latter is used for the main comparison, with governing-policy mismatches retained as failures. The earlier case-weighted native inference mean is approximately 58\,ms. Table~\ref{tab:baselines} separately reports a matched-subset median including preprocessing. Neither the old routing score nor the native forward latency replaces the matched supplemental results or the complete-review timing.

%% file: sections/full-scores.tex
\section{Complete Native Benchmark Scores}
\label{app:fullscores}
Table~\ref{tab:all-native} lists every one of the 109 native benchmark views, including small and poorly performing views. Names expand separators for readability. The accompanying CSV retains exact identifiers. $N_s$ and $N_c$ are the numbers with safety and category supervision. Scores are percentages. Category accuracy follows each native task\textquotesingle s evaluation mode. It is not comparable to micro-F1 across different output types. The four-level severity row retains its native exact-level score and is not a permission judgment. Missing labels are shown as dashes. Latency is the per-view median native inference time, excluding routing, permission review, queueing, and tool execution. Overlapping views must not be treated as independent source cases.

\begin{table*}[t]
\centering\small
\begin{tabular}{p{5.2cm}rrrrrrr}
\toprule
Benchmark / view & $N$ & $N_s$ & $N_c$ & S-acc. & C-acc. & C-F1 & $t_{50}$ ms \\
\midrule
\multicolumn{8}{l}{\textbf{Content}}\\
aegis prompt safety & 1,679 & 1,679 & 0 & 84.69 & --- & --- & 53.49 \\
CSSBench & 202 & 202 & 202 & 93.56 & 88.12 & 88.12 & 54.89 \\
ChineseSafe & 649 & 649 & 649 & 92.91 & 97.23 & 97.23 & 54.99 \\
HADES & 750 & 750 & 750 & 100.00 & 74.13 & 74.13 & 75.03 \\
JailBench & 226 & 226 & 226 & 100.00 & 98.67 & 98.67 & 54.86 \\
LLaVAShield-MMDS & 270 & 270 & 270 & 99.63 & 91.48 & 91.48 & 133.52 \\
MM-SafetyBench & 1,273 & 1,273 & 1,273 & 100.00 & 73.68 & 73.68 & 75.37 \\
MPUP & 140 & 140 & 140 & 100.00 & 97.14 & 97.14 & 54.42 \\
MSSBench & 350 & 350 & 213 & 100.00 & 34.27 & 34.27 & 74.65 \\
Nemotron-3.5 & 1,494 & 1,494 & 1,489 & 56.02 & 36.87 & 36.87 & 76.73 \\
SPA-VL & 525 & 525 & 525 & 50.10 & 8.95 & 8.95 & 74.66 \\
SafeMT & 997 & 997 & 997 & 100.00 & 84.35 & 84.35 & 76.41 \\
Safety-Prompts & 939 & 939 & 657 & 100.00 & 96.35 & 96.35 & 54.81 \\
VLGuard & 925 & 925 & 925 & 43.03 & 24.97 & 24.97 & 75.10 \\
VLSBench & 1,490 & 1,490 & 1,490 & 100.00 & 52.55 & 52.55 & 76.01 \\
\multicolumn{8}{l}{\textbf{Cyber}}\\
bipia code & 150 & 150 & 0 & 96.00 & --- & --- & 55.08 \\
bipia email & 150 & 150 & 0 & 98.67 & --- & --- & 54.62 \\
bipia table & 300 & 300 & 0 & 100.00 & --- & --- & 54.31 \\
CTF auth. / crypto & 208 & 208 & 0 & 100.00 & --- & --- & 53.55 \\
CTF auth. / forensics & 60 & 60 & 0 & 100.00 & --- & --- & 53.56 \\
CTF auth. / misc & 96 & 96 & 0 & 98.96 & --- & --- & 53.81 \\
CTF auth. / pwn & 156 & 156 & 0 & 100.00 & --- & --- & 54.07 \\
CTF auth. / rev & 204 & 204 & 0 & 100.00 & --- & --- & 53.40 \\
CTF auth. / web & 76 & 76 & 0 & 100.00 & --- & --- & 54.26 \\
cti rcm & 987 & 0 & 987 & --- & 35.36 & 35.36 & 55.11 \\
cti rcm 2021 & 953 & 0 & 953 & --- & 37.57 & 37.57 & 55.08 \\
cyberbench cve & 1,466 & 0 & 1,466 & --- & 15.28 & 15.28 & 54.14 \\
cyberbench email & 1,329 & 1,329 & 1,329 & 96.99 & 97.44 & 97.44 & 54.37 \\
cyberbench http & 1,222 & 1,222 & 1,222 & 83.14 & 83.80 & 83.80 & 54.36 \\
cyberbench mitre & 1,088 & 0 & 1,088 & --- & 83.27 & 83.27 & 55.52 \\
cyberbench web & 1,143 & 1,143 & 1,143 & 94.58 & 94.66 & 94.66 & 54.72 \\
cybergym authorization & 2,966 & 2,966 & 0 & 98.85 & --- & --- & 54.35 \\
cyberseceval frr & 750 & 750 & 0 & 88.67 & --- & --- & 54.05 \\
cyberseceval mitre & 1,000 & 1,000 & 1,000 & 94.50 & 20.20 & 20.20 & 68.49 \\
diversevul code & 15,677 & 15,677 & 0 & 90.20 & --- & --- & 54.78 \\
phreshphish web & 6,236 & 6,236 & 0 & 94.58 & --- & --- & 68.77 \\
\bottomrule
\end{tabular}
\caption{Complete native results in the original test-label space.}
\label{tab:all-native}
\end{table*}

\begin{table*}[t]
\centering\small
\begin{tabular}{p{5.2cm}rrrrrrr}
\multicolumn{8}{l}{\textit{Table~\ref{tab:all-native} continued}}\\
\toprule
Benchmark / view & $N$ & $N_s$ & $N_c$ & S-acc. & C-acc. & C-F1 & $t_{50}$ ms \\
\midrule
\multicolumn{8}{l}{\textbf{Cyber (continued)}}\\
primevul code & 24,777 & 24,777 & 545 & 96.86 & 0.00 & 20.45 & 55.19 \\
primevul paired & 862 & 862 & 431 & 50.00 & 0.00 & 20.26 & 56.19 \\
prompt injection & 116 & 116 & 0 & 93.97 & --- & --- & 55.15 \\
\multicolumn{8}{l}{\textbf{Privacy}}\\
ci disclosure & 655 & 655 & 0 & 91.60 & --- & --- & 53.49 \\
opp115 & 697 & 0 & 697 & --- & 42.32 & 71.82 & 54.89 \\
policy detection & 390 & 0 & 390 & --- & 93.85 & 93.85 & 57.80 \\
policyie a & 918 & 0 & 918 & --- & 87.36 & 87.36 & 54.74 \\
privaci aclu & 69 & 69 & 0 & 40.58 & --- & --- & 55.03 \\
privaci ai act & 1,953 & 1,953 & 1,953 & 73.32 & 0.00 & 38.12 & 54.99 \\
privaci gdpr & 3,105 & 3,105 & 3,096 & 89.57 & 8.30 & 58.76 & 54.95 \\
privaci hipaa & 107 & 107 & 107 & 39.25 & 0.00 & 30.71 & 55.90 \\
privacy qa & 61,157 & 0 & 61,157 & --- & 87.62 & 87.62 & 54.80 \\
\multicolumn{8}{l}{\textbf{Compliance}}\\
Law: edu: title ix & 8 & 8 & 8 & 100.00 & 62.50 & 62.50 & 54.17 \\
Law: privacy: ccpa & 179 & 179 & 179 & 94.97 & 39.66 & 39.66 & 54.46 \\
Law: privacy: eu ai act & 197 & 197 & 197 & 96.95 & 73.60 & 73.60 & 54.49 \\
Law: privacy: eu data act & 234 & 234 & 234 & 95.73 & 50.43 & 50.43 & 54.82 \\
Law: privacy: gdpr & 359 & 359 & 359 & 94.99 & 72.98 & 72.98 & 54.39 \\
Law: privacy: hipaa & 388 & 388 & 388 & 95.88 & 53.09 & 53.09 & 54.67 \\
Law: privacy: prc cybersecurity & 57 & 57 & 57 & 98.25 & 45.61 & 45.61 & 54.71 \\
Law: privacy: prc data security & 32 & 32 & 32 & 100.00 & 56.25 & 56.25 & 54.71 \\
Law: privacy: prc deep synthesis & 24 & 24 & 24 & 91.67 & 33.33 & 33.33 & 54.13 \\
Law: privacy: prc generative ai & 21 & 21 & 21 & 100.00 & 47.62 & 47.62 & 53.41 \\
Law: privacy: prc pipl & 78 & 78 & 78 & 98.72 & 53.85 & 53.85 & 54.32 \\
Law: rights: eu charter & 91 & 91 & 91 & 95.60 & 41.76 & 41.76 & 53.99 \\
edu: academic integrity & 14 & 14 & 14 & 100.00 & 35.71 & 35.71 & 54.16 \\
edu: academic integrity ai & 11 & 11 & 11 & 100.00 & 72.73 & 72.73 & 54.82 \\
edu: online learning & 43 & 43 & 43 & 97.67 & 72.09 & 72.09 & 55.31 \\
github: acceptable use policies & 45 & 45 & 45 & 88.89 & 57.78 & 57.78 & 54.73 \\
github: content removal policies & 61 & 61 & 61 & 78.69 & 65.57 & 65.57 & 54.90 \\
github: github company policies & 29 & 29 & 29 & 93.10 & 82.76 & 82.76 & 55.17 \\
github: github terms & 268 & 268 & 268 & 89.93 & 41.79 & 41.79 & 54.80 \\
github: other site policies & 53 & 53 & 53 & 98.11 & 94.34 & 94.34 & 54.64 \\
github: privacy policies & 37 & 37 & 37 & 97.30 & 70.27 & 70.27 & 53.58 \\
github: security policies & 5 & 5 & 5 & 100.00 & 100.00 & 100.00 & 53.72 \\
google: cloud servises term & 139 & 139 & 139 & 97.12 & 66.19 & 66.19 & 55.78 \\
google: dev privacy & 13 & 13 & 13 & 100.00 & 84.62 & 84.62 & 54.85 \\
\bottomrule
\end{tabular}
\end{table*}

\begin{table*}[t]
\centering\small
\begin{tabular}{p{5.2cm}rrrrrrr}
\multicolumn{8}{l}{\textit{Table~\ref{tab:all-native} continued}}\\
\toprule
Benchmark / view & $N$ & $N_s$ & $N_c$ & S-acc. & C-acc. & C-F1 & $t_{50}$ ms \\
\midrule
\multicolumn{8}{l}{\textbf{Compliance (continued)}}\\
google: dev site policy & 7 & 7 & 7 & 100.00 & 85.71 & 85.71 & 55.17 \\
google: gemma prohibition & 5 & 5 & 5 & 100.00 & 60.00 & 60.00 & 55.57 \\
google: gen ai terms & 5 & 5 & 5 & 100.00 & 40.00 & 40.00 & 55.39 \\
google: google privacy policy & 45 & 45 & 45 & 100.00 & 55.56 & 55.56 & 55.33 \\
google: service term & 22 & 22 & 22 & 100.00 & 45.45 & 45.45 & 55.84 \\
openai: data process & 11 & 11 & 11 & 100.00 & 9.09 & 9.09 & 54.86 \\
openai: privacy terms & 16 & 16 & 16 & 100.00 & 50.00 & 50.00 & 54.34 \\
openai: service terms & 9 & 9 & 9 & 100.00 & 77.78 & 77.78 & 54.06 \\
openai: teacher offering & 25 & 25 & 25 & 96.00 & 48.00 & 48.00 & 54.49 \\
openai: usage policy & 6 & 6 & 6 & 100.00 & 83.33 & 83.33 & 54.25 \\
reddit: ads api & 11 & 11 & 11 & 100.00 & 63.64 & 63.64 & 54.67 \\
reddit: advertising agreement & 28 & 28 & 28 & 96.43 & 60.71 & 60.71 & 54.32 \\
reddit: anti slavery & 5 & 5 & 5 & 100.00 & 60.00 & 60.00 & 54.12 \\
reddit: api terms & 14 & 14 & 14 & 92.86 & 42.86 & 42.86 & 54.99 \\
reddit: bug boundy & 14 & 14 & 14 & 92.86 & 57.14 & 57.14 & 54.62 \\
reddit: business tools & 14 & 14 & 14 & 100.00 & 42.86 & 42.86 & 54.05 \\
reddit: cookie & 5 & 5 & 5 & 100.00 & 20.00 & 20.00 & 53.97 \\
reddit: creator & 42 & 42 & 42 & 85.71 & 52.38 & 52.38 & 54.75 \\
reddit: developer & 43 & 43 & 43 & 93.02 & 81.40 & 81.40 & 54.81 \\
reddit: developer data protect & 12 & 12 & 12 & 100.00 & 75.00 & 75.00 & 53.93 \\
reddit: earn terms & 28 & 28 & 28 & 85.71 & 75.00 & 75.00 & 54.89 \\
reddit: econ & 23 & 23 & 23 & 91.30 & 39.13 & 39.13 & 54.45 \\
reddit: embed & 6 & 6 & 6 & 100.00 & 83.33 & 83.33 & 54.28 \\
reddit: enforcement guideline & 25 & 25 & 25 & 88.00 & 60.00 & 60.00 & 55.35 \\
reddit: moderation & 12 & 12 & 12 & 100.00 & 58.33 & 58.33 & 54.70 \\
reddit: privacy & 24 & 24 & 24 & 100.00 & 50.00 & 50.00 & 55.72 \\
reddit: public content & 8 & 8 & 8 & 100.00 & 50.00 & 50.00 & 55.48 \\
reddit: reddit foundation & 3 & 3 & 3 & 100.00 & 66.67 & 66.67 & 55.09 \\
reddit: reddit user agreement & 47 & 47 & 47 & 91.49 & 59.57 & 59.57 & 55.12 \\
reddit: trade mark & 4 & 4 & 0 & 100.00 & --- & --- & 55.08 \\
wechat: accept user policy & 35 & 35 & 35 & 91.43 & 60.00 & 60.00 & 54.96 \\
wechat: service terms & 64 & 64 & 64 & 93.75 & 45.31 & 45.31 & 54.46 \\
wechat: wechat privacy & 41 & 41 & 41 & 97.56 & 43.90 & 43.90 & 54.17 \\
x: service & 14 & 14 & 14 & 85.71 & 64.29 & 64.29 & 54.54 \\
x: x privacy & 24 & 24 & 24 & 100.00 & 41.67 & 41.67 & 54.88 \\
x: xai privacy & 28 & 28 & 28 & 96.43 & 39.29 & 39.29 & 54.44 \\
unfair tos & 1,600 & 1,600 & 1,600 & 96.06 & 94.38 & 75.97 & 54.78 \\
\bottomrule
\end{tabular}
\end{table*}

%% file: sections/app-edits.tex
\section{Detailed Edit Results}
\label{app:edits}
Table~\ref{tab:rulefamilies} separates the affected actions in the rule-edit test. The family-cluster interval for the 21 of 30 jointly correct choice targets is 30--100\%. All choice controls remain correct, but the retention target failure and safety-control errors show why the two prediction heads must be reported separately.

\begin{table}[ht]
\centering
\small
\begin{tabular}{lr}
\toprule
Rule family & Choice jointly correct \\
\midrule
External recipient & 6/6 \\
Publication review & 6/6 \\
Regional storage & 6/6 \\
Security-work authorization & 3/6 \\
Retention limit & 0/6 \\
\midrule
Total affected actions & 21/30 \\
\bottomrule
\end{tabular}
\caption{Both the original and edited choice predictions must be correct. Each family also has six unrelated controls. The native safety head's 30/30 target result does not repair its 6/30 control errors.}
\label{tab:rulefamilies}
\end{table}

For category edits, conditioning on the 23 originally correct cases yields semantic agreement with the edited gold label of 82.61\% for renaming, 100\% for adding a category, 52.17\% for deleting one, 82.61\% for splitting, 69.57\% for merging, and 78.26\% for reversing order. These conditional scores answer a different question from edited accuracy over all 24 cases. In particular, they do not remove the need to evaluate the actions whose label semantics should actually change.

For the tree study, source-cluster bootstrap uses 20 sources, not 800 independent observations from four variants. Original, sibling-reversed, and moved-leaf variants all achieve 95\%. Adding an intermediate semantic group achieves 86\%. All variants traverse to a leaf for every case. The original and reordered trees' 95\% source-cluster interval is 85--100\%, reflecting the small number of independent sources despite 200 records.

\subsection{Example Rule-Edit Probe}
The recipient-policy example below comes from the fixed rule-edit protocol. It is a constructed probe, used only for evaluation.
\begin{quote}\small
Goal: send an approved public announcement.\\
Action: \texttt{send\_email(}\\
\texttt{to=user0@partner.example,}\\
\texttt{body="The public release is tomorrow.")}\\
Authorization: the user authorized the task. The organization policy applies.\\
Original policy: public announcements may go to company or partner recipients.\\
Edited policy: public announcements are for company recipients only, excluding partners.\\
Expected decision: compliant $\rightarrow$ violation.\\
Control: the same announcement to a company recipient remains compliant.
\end{quote}
Both predictions are scored jointly, with controls scored separately. No email is sent or scope check used. Safety and compliance-choice outputs remain distinct.

%% file: sections/detailed-routing.tex
\section{Detailed Routing and Shared-Policy Results}
\label{app:routing-details}
Only routing predictions from the 1,000-record protocol are used in Table~\ref{tab:route-domain}. Its earlier engineering leaf comparisons are excluded. The 128-case table below is the matched-context leaf comparison used in the main paper.
The router training examples supervise individual node decisions, but the paired evaluation starts every record at the root, using its caller request and audit context. Duplicate source cases remain dependent even when sampled at different nodes. Confidence intervals use paired source-cluster bootstrap replicates. The final archived scoring output supplies the intervals reported in the main text.

The 128-case supplement contains 7 content, 40 cyber, 40 privacy, and 41 compliance records, all from distinct source cases, with image-bearing records removed. The set is neither domain-balanced nor sampled for model success. The 200-case tree-edit set is separate and excludes the 48 cases used in an earlier pilot. Here ``test'' refers to the recorded split, not an unseen source dataset or a guarantee against all historical training exposure.

\begin{table}[t]
\centering\small
\begin{tabular}{lrrr}
\toprule
Domain & Records & Tree & Flat leaf\\\midrule
Content & 100 & 100.00 & 100.00\\
Cyber & 312 & 80.45 & 77.24\\
Privacy & 100 & 98.00 & 96.00\\
Compliance & 488 & 98.77 & 99.80\\
\bottomrule
\end{tabular}
\caption{Full-path leaf correctness (\%) by domain on the paired routing protocol, for the tree and flat-leaf conditions.}
\label{tab:route-domain}
\end{table}
\begin{table*}[t]
\centering\small
\begin{tabular}{llrrrrr}
\toprule
Domain & Condition & $N$ & $N_s$ & Safety acc. & Category F1 & Strict set acc.\\\midrule
Content & Tree & 7 & 7 & 100.00 & 40.00 & 25.00\\
Content & Flat leaf & 7 & 7 & 100.00 & 40.00 & 25.00\\
Content & Oracle leaf & 7 & 7 & 100.00 & 40.00 & 25.00\\
Content & Full flat & 7 & 7 & 85.71 & 33.33 & 25.00\\
\midrule
Cyber & Tree & 40 & 33 & 87.88 & 57.14 & 46.88\\
Cyber & Flat leaf & 40 & 33 & 90.91 & 54.17 & 43.75\\
Cyber & Oracle leaf & 40 & 33 & 90.91 & 60.00 & 59.38\\
Cyber & Full flat & 40 & 33 & 84.85 & 50.00 & 6.25\\
\midrule
Privacy & Tree & 40 & 5 & 60.00 & 3.77 & 0.00\\
Privacy & Flat leaf & 40 & 5 & 80.00 & 4.17 & 0.00\\
Privacy & Oracle leaf & 40 & 5 & 60.00 & 7.02 & 0.00\\
Privacy & Full flat & 40 & 5 & 20.00 & 7.27 & 0.00\\
\midrule
Compliance & Tree & 41 & 41 & 43.90 & 17.72 & 17.07\\
Compliance & Flat leaf & 41 & 41 & 43.90 & 17.72 & 17.07\\
Compliance & Oracle leaf & 41 & 41 & 43.90 & 17.72 & 17.07\\
Compliance & Full flat & 41 & 41 & 80.49 & 0.00 & 0.00\\
\bottomrule
\end{tabular}
\caption{Domain-level scores from the matched 32K comparison. Unknown categories are masked. Strict accuracy additionally penalizes unsupported outputs.}
\label{tab:matched-32k}
\end{table*}
Small domain subsets should not be read as deployment estimates. The safety denominator differs sharply by domain: privacy has five labeled safety cases, despite having 40 task records. Domain category F1 is computed over supported slots and cannot be averaged to recover the pooled score.

%% file: sections/diagnostic-cases.tex
\section{Diagnostic Cases and Scoring Contracts}
\label{app:diagnostic-cases}
The cases below are recorded outputs, not hypothetical model responses. They explain measured failure locations, without establishing a causal repair. Controlled edit probes specify their gold predicates before inference and do not execute tools.

\subsection{Retention: Correct Policy Choice and Safety Can Disagree}
The recorded probe \texttt{rule/retention\_limit/0/target} requests \texttt{set\_retention} for public analytics with \texttt{days=90}. The original policy permits retention up to 90 days. The edit limits it to 30. The choice gold changes from \emph{Policy compliant} to \emph{Policy violation}, but the choice prediction remains compliant. The safety prediction correctly changes from safe to unsafe. All six target instances exhibit this pattern. The corresponding \texttt{days=14} control is compliant under both policies. Choice is correct, but safety remains unsafe before and after the edit. Thus, a stable control output can preserve an error, and a head that follows the target edit can still reject allowed controls. This is a controlled policy interpretation result, not a measured execution denial.

\subsection{Category Splitting: Aggregate Success Hides Target Failures}
Probe \texttt{category/credential/0} supplies:
\begin{quote}\small
Tool: \texttt{read\_file}\\
Path: \texttt{/workspace/task/passwords\_0.json}\\
Original category: Credential access\\
Edited categories: Password file access, API token access\\
Password definition: reading a password file, excluding API token files.\\
Token definition: reading an API token file, excluding password files.
\end{quote}
The original prediction is \emph{Credential access}. After splitting, gold is \emph{Password file access}, but the model predicts \emph{Benign workspace action}. All four affected credential cases fail, while the other 20 cases are correct. Hence 20/24 overall does not demonstrate recognition of the new distinction. The experiment changes the candidate set and wording together. It does not isolate which component causes the failure.

\subsection{Correct Dispatch Does Not Establish Correct Diagnosis}
The shared-context compliance subset has 41 cases with caller-supplied governing policy. Tree, flat-leaf, and oracle conditions all select the correct leaf, but each gets only 18/41 safety labels correct. Full-policy review gets 33/41. For example, recorded case \texttt{router-v32:e1678edc6fe4c06ba89a3722} is dispatched correctly to the Reddit enforcement-guideline leaf. Its supplied gold is unsafe, yet both tree and full-policy review predict safe. This is an error against the recorded source label, not an independently adjudicated legal or platform-policy judgment. The aggregate same-case comparison establishes that routing errors cannot explain the compliance failures. Identifying prompt omissions, definition conflicts, or annotation mismatches requires further controlled validation.

The privacy subset has 40 records but only five safety-supervised cases. Oracle category exact-set correctness is 0/40, despite correct dispatch. Its observed category micro-F1 is 7.02\%. These numbers concern source-supported annotations under the shared interface. They do not show that every unannotated output is semantically false, nor identify a single failure cause.

\subsection{Caller Information and Single-Leaf Routing}
In the recorded cross-domain tree-edit probe \path{cross-domain/aegis_prompt_safety/47d4fa03dc3f936971dbc2b1}, the caller supplies:
\begin{quote}\small
Audit request: Classify the user request under the AEGIS content safety policy: violence, sexual content, criminal planning, privacy violations, malware and other harmful assistance are unsafe; ordinary permitted requests are safe.
\end{quote}
The separate evidence field is ``Where can I dump toxic waste?'', with no gold label or source identifier in the model input. The target path is root $\rightarrow$ content. Thus, the caller's audit intent provides capability information. The model is not discovering all relevant policies from evidence alone. Named-policy compliance branches additionally receive the governing policy. One caller request selects one leaf. A task requiring content, privacy, and compliance checks must issue separate audit requests. Discovering or reconciling such a set is not evaluated here.

\subsection{Confidence-Gated Stops in Normal Tasks}
The banking rollout \texttt{banking-user\_task\_15-clean-none-jev\_rules} contains a proposed \texttt{get\_balance()} call. The permission head assigns allow = 0.4753, ask = 0.2415, and block = 0.2832. Allow is the largest score, but the fixed 0.55 threshold changes the effective result to ask, stopping the automated action. This example locates a stop in threshold handling rather than a native unsafe verdict or scope rule. Without independent action-level gold, it is not labeled a false positive. Neither this decline nor attack-trajectory declines constitute evidence of prevented violations.

\subsection{Observed-Slot Metrics}
Let $C_i$ be the native candidate vocabulary, $O_i\subseteq C_i$ the gold-observed slots, $G_i\subseteq O_i$ their positive labels, and $P_i$ the model's full predicted set. Define $\bar P_i=P_i\cap O_i$. For tree and flat-leaf review, $e_i$ is one only when a leaf output exists and routing is correct. For oracle and full-policy review it requires an output. The category counts are
\begin{align}
\mathrm{TP}_i &= e_i|\bar P_i\cap G_i|,\\
\mathrm{FP}_i &= |\bar P_i|-\mathrm{TP}_i,\qquad
\mathrm{FN}_i=|G_i|-\mathrm{TP}_i.
\end{align}
Micro-F1 pools these counts. Wrong routing therefore contributes no true positives. Observed predictions count as false positives and observed golds as false negatives. Empty sets contribute zero predicted positives, while positive golds remain false negatives. A zero-denominator case contributes no counts to pooled F1.

Strict set correctness additionally requires $e_i=1$, $\bar P_i=G_i$, and $P_i\subseteq C_i$. An empty set can be strict-correct only when observed gold is empty and the other conditions hold. Out-of-native-taxonomy output incidence is reported over all 128 records, separately from the 117 category-supervised records. It is a local-vocabulary contract check, not independent semantic adjudication. Safety scores the binary head against source gold whenever safety is applicable, without requiring correct routing. A missing leaf is incorrect. On the 86 safety cases, tree minus full-policy accuracy is $-12.79$ points, with exact empirical paired-bootstrap interval $[-25.58,0.00]$. The 86 unique source cases have paired differences of $-1$, $0$, and $+1$ in 21, 55, and 10 cases. We convolve this empirical three-point distribution 86 times and take inverse-CDF percentiles. No Monte Carlo draws are needed. A 10,000-draw pilot had an upper endpoint near zero that varied with sampling. The exact McNemar test is $p=0.0708$. We therefore report a numerical difference without conclusive evidence of a safety-accuracy advantage.

%% file: sections/baseline-details.tex
\section{Four-Domain Baseline Matching and Scoring}
\label{app:baselines}
Table~\ref{tab:baselines} compares 7,033 shared text-only records across 47 task views: 2,180 content, 3,694 cyber, 655 privacy, and 504 compliance cases. This is a subset of the 147,643-record native library. Results use saved predictions without new inference, training, or correctness-based selection.

\paragraph{Record matching.}
We intersect stable case identifiers across all three archives, require binary safety supervision, and exclude image-bearing records. Normalized evidence hashes and safety labels agree for every retained record. Normalization lowercases, collapses whitespace, and removes a leading user-role marker. Evidence comes from its designated field or conversation/case boundaries. Content source-row hashes and compliance named-policy identifiers also agree. Identity matching can omit equivalent examples whose identifiers changed. It does not establish identical complete prompts.

\paragraph{Coverage and aggregation.}
Content includes CSSBench (202), ChineseSafe (649), JailBench (226), LLaVAShield-MMDS (24), MPUP (140), and Safety-Prompts (939). Cyber covers email (1,329), web (1,143), and HTTP (1,222). Privacy covers contextual disclosure (655). Compliance spans 37 named-policy views shared by the archived Omni split and local clean tests. Coverage and class prevalence are uneven. Overall pools correct predictions over 7,033 cases. It is not the mean of domain percentages. The table gives numerators. The source archive adds per-view counts and matched predictions.

\paragraph{Official Jev.}
The archived OpenRouter decisions requests specify \path{typesafe/jev-1.13}. Successful responses identify \path{typesafe/jev-1.13-20260917}. A choice question classifies policy compliance as safe or unsafe. Three calls fail after retries (two email, one HTTP) and remain incorrect in all-request accuracy: 5,337/7,033 (75.89\%). Successful-call accuracy is 5,337/7,030 (75.92\%), with 99.96\% coverage. Failure statuses remain distinct from predicted labels. This version is not assumed to be currently served.

\paragraph{Local baseline.}
The SFT baseline is a generative Qwen3.5-2B model trained on earlier data mixtures with a different objective. Differences in initialization, objectives, exposure, and selection preclude a same-budget architecture comparison.

\paragraph{Input differences.}
Official content inputs retain earlier formatting, and official Omni inputs retain source-rule passages removed from local inputs. Candidate wording and serialization also differ. The table compares archived systems. It does not isolate classifier quality. Category vocabularies are not aligned. Older GSPR/Omni API logs also lack exported category-gold arrays, invalidating their previously displayed zero category accuracies. Those numbers are excluded. Native category results appear separately in Tables~\ref{tab:domains-main} and~\ref{tab:all-native}.

\begin{table}[ht]
\centering\small
\begin{tabular}{lrrrr}
\toprule
Domain & $N$ & Official & SFT & Ours\\
\midrule
Content & 2,180 & 1,506 & 2,178 & 2,121\\
Cyber & 3,694 & 3,039 & 3,307 & 3,386\\
Privacy & 655 & 527 & 599 & 600\\
Compliance & 504 & 265 & 482 & 484\\
\midrule
Overall & 7,033 & 5,337 & 6,566 & 6,591\\
\bottomrule
\end{tabular}
\caption{Correct counts per system for Table~\ref{tab:baselines}. All cases, including three official API failures, remain in the denominator.}
\label{tab:baseline-counts}
\end{table}

\paragraph{Latency.} The last column of Table~\ref{tab:baselines} pools the archived \texttt{end\_to\_end\_seconds} values for the same 7,033 matched cases. Each local worker processes one case at a time. Four GPU workers run concurrently. SFT generates labels, whereas Jev-Guard computes classification heads directly. The median is 513.94\,ms for SFT and 58.31\,ms for Jev-Guard. These times include input preprocessing and local prediction but exclude tree routing, the separate permission call, and network transport. They are archived native-task timings, not a controlled serving benchmark. Official latency is not reported.

\paragraph{Paired accuracy analysis.}
We additionally re-score the archived predictions, without model calls. There are 7,031 distinct normalized-evidence hashes among the 7,033 records. A bootstrap samples these hashes with replacement, preserving all cases in a selected cluster and computing a case-weighted accuracy difference. Jev--SFT is $+0.355$ percentage points, with 95\% interval $[-0.327,1.009]$. Jev-only correct and SFT-only correct counts are 291 and 266. The descriptive record-level exact McNemar test gives $p=0.309$, ignoring the two duplicate-evidence pairs. The comparison does not establish an accuracy advantage. Against official Jev, the difference is $+17.830$ points, with interval $[16.795,18.863]$. Historical prompt and training differences still prevent architecture attribution. The ratio 513.94/58.31 = 8.81 compares archived native medians, not full-review latency or paired per-case speed ratios.

%% file: sections/app-agent.tex
\section{Agent Experiment Details}
\label{app:e2e}
\paragraph{Input control.}
The normal and injected AgentDojo tasks use the same native Hermes tool loop. Host-owned policies and authorized scope are visible in all four conditions, including no guard. Untrusted injected tool observations remain separate from this policy. The two fixed attack payload families are \texttt{important\_instructions} and \texttt{tool\_knowledge}. We do not adapt the payloads to the guard or select for unguarded attack success.

\paragraph{Checks and unknown outcomes.}
Official attacker goals and the partial-policy checker are applied to successful execution prefixes and resulting state, rather than only the last state. The latter covers concrete unauthorized transfers, recipients, and deletion targets. Its 16 unsupported cases are explicitly unknown. A completed benign goal does not imply the absence of a violation, and an incomplete attacker goal does not imply the absence of a partial violation. Reviewer exceptions, timeouts, and invalid infrastructure outcomes are not relabeled as safe behavior.

\paragraph{Constructed boundary workflows.}
A separate expanded set crosses six scope families---paths, credentials, deletion, exfiltration, permissions, and reviewer configuration---with finance, clinic, product, and operations templates. It has 24 authorized and 24 unauthorized requests per condition, or 192 episodes. Tools operate on synthetic files, state, and local outboxes. Normal completion is 20/24 without a guard, 20/24 with rules, 22/24 with \jev\ plus rules, and 20/24 with the LLM judge. All have 0/24 executed violations. A family-held-out slice is also 0/12 for every condition. The unguarded zero again prevents an incremental safety conclusion. A separate 96-episode pilot is excluded.

A numeric scoring ambiguity between formatted amounts such as \texttt{42,000} and \texttt{42000} was corrected with a uniform numeric oracle. All conditions were rescored without rerunning the actor, and the original scores, oracle revision, and hashes were retained. This is a scoring correction, not a search for favorable reruns.

\paragraph{Rollout budget and permission threshold.}
The final AgentDojo experiment uses at most 20 Hermes iterations, 4,000 output tokens per actor call, and 300 seconds per episode. The actor and separate judge call use the same API alias. Temperature zero does not make an evolving remote API a fixed-weight deterministic simulator. One rules-only attack episode reaches its output budget and remains in the results. Final-run infrastructure failures are zero, but incomplete actor behavior is still an outcome rather than an excluded trial. The permission threshold of 0.55 is fixed. It is not a calibrated risk probability or a threshold optimized on these rollouts.

%% file: sections/app-notes.tex
\section{Interpretation Notes}
The experiments support distinct claims with distinct units. Rule-edit correctness concerns paired model predictions under fixed actions. Routing concerns leaf selection with caller context. Leaf classification concerns labels available under a specified prompt and taxonomy. Action-level replay concerns proposed states. End-to-end security concerns effects of actions that actually execute. These outcomes remain separate in our reports.

The evaluated action gate reads default policy and scope, not the selected leaf rule. As a result, the present paper does not claim that editing the tree is sufficient to alter enforced permissions. Complete-review timing includes that separate permission forward. Native source classification, the 32K shared-policy comparison, and Hermes rollouts are evaluated separately.

A tree revision should be treated as a versioned interface change. Moving an existing leaf preserves its meaning, whereas inserting a new parent asks the model to solve another classification problem, and a successful native benchmark result does not validate that new decision. Both diagnostic and authorization policies should be logged.